\documentclass[a4paper,fleqn]{cas-sc}
   
\AddToHook{begindocument/before}{\ifdefined\bibsep\else\newlength\bibsep\fi}

\def\tsc#1{\csdef{#1}{\textsc{\lowercase{#1}}\xspace}}        
\tsc{WGM}
\tsc{QE}
\tsc{EP}
\tsc{PMS}
\tsc{BEC}
\tsc{DE}     
\usepackage[numbers]{natbib}
\usepackage{graphicx}
\usepackage{color}
\usepackage{dcolumn}
\usepackage{epsfig}
\usepackage{amsfonts}
\usepackage{array}
\usepackage{microtype}
\usepackage{multirow}
\usepackage{adjustbox}
\usepackage[english]{babel}
\usepackage{epstopdf}
\usepackage{blindtext}
\usepackage{subcaption}
\usepackage{bm}
\usepackage{amsmath}
\usepackage{amssymb}
\usepackage{lineno,hyperref}
\usepackage{float}
\usepackage{duckuments}
\def \g{\gamma}

\def \be{\begin{equation}}
\def \ee{\end{equation}}
\def \ben{\begin{eqnarray}}
\def \een{\end{eqnarray}}

\newcommand{\orcid}[1]{\href{https://orcid.org/#1}{\includegraphics[height=\fontcharht\font`\B]{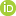}}}
\usepackage[utf8]{inputenc}
\usepackage[T1]{fontenc}

\begin{document}
   
\let\WriteBookmarks\relax
\def\floatpagepagefraction{1}
\def\textpagefraction{.001}                  

\title[mode = title]{A cosmographic analysis using DESI- DR2 and strong lensing: II. Distance Ratio measurements }   

\author[1]{Darshan Kumar}
\ead{kumardarshan@hnas.ac.cn}
\address[1]{Institute for Gravitational Wave Astronomy, Henan Academy of Sciences, Zhengzhou, Henan 450046, China}
% \orcid{0000-0001-6665-8284}
 
% Second Author
\author[2]{Deepak Jain}     
\ead{djain@ddu.du.ac.in}
\address[2]{Deen Dayal Upadhyaya College, University of Delhi, Dwarka, New Delhi 110078, India}

\author[3]{Shobhit Mahajan}
\ead{sm@physics.du.ac.in}
\address[3]{Department of Physics and Astrophysics, University of Delhi, Delhi 110007, India}

\begin{abstract}
    The distance ratios derived from strong lensing systems, combined with complementary cosmological observations, allow for the study of cosmic expansion and curvature without assuming a fixed background cosmological model. In this work, we perform an analysis of cosmic expansion using the latest Type Ia supernova samples, including PantheonPlus, Union3, and DES Y5, combined with baryon acoustic oscillation dataset from DESI DR2 and strong-lensing distance ratios. The cosmic expansion is carried out to fourth order in the variable $y = z/(1+z)$, which allows constraints on the present-day deceleration, jerk, and snap parameters $(q_0, j_0, s_0)$. The analysis utilizes the distance sum rule to provide an independent determination of the spatial curvature parameter $\Omega_{k0}$ without assuming any specific cosmological dynamics. Our results from combining strong lensing distance ratios with each supernova dataset indicate that a flat Universe remains consistent at the 95\% confidence level, and the inclusion of DESI-DR2 measurements tightens the parameter intervals while preserving agreement with flat geometry at the 68\% confidence level, in line with standard cosmology. The inferred values of $q_0$ and $j_0$ are compatible with $\Lambda$CDM predictions for all dataset combinations. The constraints on $s_0$ remain weak, although modest improvement appears after DESI DR2 data are included. This work represents the second and final paper in the two-part cosmography study.
\end{abstract}
    
\maketitle

% \section{Introduction}\label{sec_intro}    
\section{Introduction}\label{sec_intro}     

The standard spatially flat $\Lambda$CDM model, which incorporates a cosmological constant and cold dark matter, successfully explains a wide range of cosmological observations. Nevertheless, unresolved tensions such as the Hubble constant discrepancy \cite{2020A&A...641A...6P,2022ApJ...934L...7R}, inconsistencies in the spatial curvature parameter \cite{2020NatAs...4..196D,2021PhRvD.103d1301H}, and theoretical issues like fine-tuning and coincidence problems \cite{1989RvMP...61....1W,1999PhRvL..82..896Z,2000astro.ph..5265W} persist. These challenges call for model-independent approaches that avoid assumptions tied to specific cosmologies.        

The distance sum rule (DSR) has emerged as an effective tool to test spatial curvature without relying on background cosmological models, provided the universe adheres to the FLRW metric \cite{2015PhRvL.115j1301R}. %Any violation of the rule signals either data inconsistencies or deviations from the standard metric framework.   
Previous works have applied the DSR with observations such as Type Ia supernovae, gamma-ray bursts, and gravitational-wave standard sirens \cite{2019PhRvL.123w1101C,2023MNRAS.521.4963D,2019PhRvD..99h3514L}. {For more details on the broader context of model-independent tests, we refer the reader to the introduction of the first paper in this series (hereafter Paper I)} \cite{2025arXiv251100788K}.                  

% Strong gravitational lensing (SGL) has become an indispensable observational probe to test cosmological and astrophysical assumptions \cite{2010ARA&A..48...87T}. Apart from its use in time-delay cosmography, SGL also provides measurements of the distance ratio $d_A^{^{ls}}/d_A^{^{os}}$, where $d_A^{^{ls}}$ and $d_A^{^{os}}$ denote the angular diameter distances between the lens and the source, and between the observer and the source, respectively. These ratios can be used to study both the geometry of the universe and the mass distribution in lensing galaxies. The Singular Isothermal Sphere (SIS) profile, with a mass density proportional to $r^{-2}$, has been widely applied to model lens galaxies, while recent studies explore a generalized power-law profile $\rho_T(r) \propto r^{-\gamma}$ to better account for galactic structure \cite{2015ApJ...806..185C,2016RAA....16...84L,2016MNRAS.461.2192C,2017SCPMA..60h0411C}. Observations involving hundreds of SGL systems have been used to constrain cosmological parameters and lens properties under the assumption of a flat universe \cite{2019MNRAS.488.3745C,2017ApJ...834...75X}.\\

Strong gravitational lensing (SGL) systems provide a complementary observational probe that is sensitive to cosmic geometry through angular diameter distances \cite{2010ARA&A..48...87T}. Apart from their use in time-delay cosmography \cite{1964MNRAS.128..307R,2024SSRv..220...48B}, SGL also provides measurements of the distance ratio $d_A^{^{ls}}/d_A^{^{os}}$, where $d_A^{^{ls}}$ and $d_A^{^{os}}$ denote the angular diameter distances between the lens and the source, and between the observer and the source, respectively. These ratios can be used to study both the geometry of the universe and the mass distribution in lensing galaxies. Further, they can be used alongside other distance probes to constrain cosmological parameters without assuming a particular background model. 

A central observable in SGL studies is the Einstein radius $\theta_E$, which characterizes the angular size of the ring-like image formed when a source is aligned with a lens. The Einstein radius is a robust probe of the total projected mass of the lensing galaxy and, when combined with stellar kinematics, has been used to infer mass profiles and other galaxy properties \cite{2002ApJ...575...87T,2003ApJ...583..606K,2006ApJ...649..599K}. Importantly, the observed angular size depends on the distance ratio $d_A^{ls}/d_A^{os}$, making it sensitive to the cosmological model assumed. Thus, by modeling lens galaxies and comparing theoretical predictions with observed Einstein radii, one can derive constraints on cosmological parameters \cite{PhysRevD.84.023005,2010MNRAS.406.1055B,2022MNRAS.516.5187W,2017JCAP...03..028R}.

The mass distribution within lens galaxies is a key factor in such analyses. The Singular Isothermal Sphere (SIS) profile, characterized by a mass density proportional to $r^{-2}$, offers a simple yet effective model for describing lens galaxies and is widely used in lensing studies. Further investigations have explored extended power-law models (EPL), in which the total mass density and luminous mass density follow $\rho_T(r) \propto r^{-\gamma}$ and $\rho_L(r) \propto r^{-\delta}$, respectively. These models describe how the mass distribution varies with radius and allow analysis to account for differences in both luminous and dark matter components within galaxies \cite{2015ApJ...806..185C,2016RAA....16...84L,2016MNRAS.461.2192C,2017SCPMA..60h0411C,2019MNRAS.488.3745C,2017ApJ...834...75X}. Such models have been applied to study galaxy evolution, structure formation, and important cosmological parameters \cite{2019MNRAS.488.3745C,2024MNRAS.527.5311L,2024PhRvD.109b3533G}. By considering different and redshift dependent mass profiles, these models offer a more detailed description of galactic dynamics and interactions. At the same time, this approach introduces more parameters and modeling complexities, which increase uncertainties and present challenges for interpreting observational data.  
% Observations involving hundreds of SGL systems have been used to constrain cosmological parameters and lens properties under the assumption of a flat universe \cite{2019MNRAS.488.3745C,2017ApJ...834...75X}

In this analysis, we adopt the simpler SIS profile rather than extended models. Our objective is not to explore the detailed structure or evolution of lens galaxies, but rather to utilize distance ratio measurements as a clean geometrical probe to constrain cosmographic parameters and spatial curvature. By relying on the SIS model, we minimize assumptions about galaxy structure and focus on using lensing data to probe cosmic geometry in a model-independent framework.

Cosmography provides a model-independent framework to describe the kinematics of the universe’s expansion without assuming any specific form for dark energy or gravity. By expanding observables such as the scale factor or luminosity distance as a Taylor series around the present epoch, cosmography directly relates measurable quantities to parameters like the Hubble constant ($H_0$), the deceleration parameter ($q_0$), the jerk parameter ($j_0$), and higher derivatives that describe the evolution of cosmic expansion \cite{1972gcpa.book.....W,1997Sci...276...88V,2005GReGr..37.1541V}. This approach relies only on the assumption of large-scale homogeneity and isotropy, as described by the FLRW metric, and avoids dependence on a particular cosmological model or dark energy parameterization. Despite its theoretical simplicity, standard expansions in redshift suffer from convergence issues at higher redshifts, typically beyond $z \approx 1$ \cite{2004JCAP...09..009C,2004ApJ...607..665R}. To extend its applicability, alternative redshift parametrizations such as $y = z/(1+z)$ \cite{2001IJMPD..10..213C,2003PhRvL..90i1301L} and rational approximations like Padé or Chebyshev polynomials have been introduced to ensure mathematical robustness across larger redshift domains \cite{2014PhRvD..89j3506G,2012JCAP...05..024S}. Cosmography has therefore become a powerful tool to constrain cosmic expansion and curvature independently. It provides an alternative to model-dependent analyses and offers a clear path for precision cosmology in the era of high-quality observational data.

In this work, we focus on the cosmographic analysis using SGL distance ratio measurements along with other distance probes, namely Type Ia supernova compilations (PantheonPlus, DESY5, and Union3) and baryon acoustic oscillation data from DESI-DR2. We expand the cosmographic series up to the fourth order in the variable $y = z/(1+z)$ to constrain the deceleration, jerk, and snap parameters $(q_0,~j_0,~s_0)$, and we implement the DSR to probe spatial curvature $\Omega_{k0}$ independently. Compared to the first paper, which focused primarily on time-delay distances, here we emphasize the use of distance ratios as an independent geometrical probe that provides an independent constraint on cosmic expansion and geometry.   

The inclusion of SGL distance ratios in cosmographic analyses represents a novel approach. Previous studies either relied on distance ladder calibrations or assumed a flat universe, whereas this study applies the DSR framework directly to distance ratio measurements without such assumptions. By combining the largest available samples of SGL systems with supernova and BAO datasets, we offer a unique and robust way to investigate cosmic curvature and expansion. Our goal is to explore the consistency of these probes with standard cosmological predictions and to examine how their combination can enhance the reliability of cosmological constraints.

The structure of this paper is as follows. In Section \ref{sec_data_meth}, we describe the datasets and methodology used in this analysis. Section \ref{sec_resu} presents the main results, including constraints on cosmographic and curvature parameters. Finally, Section \ref{sec_disc_conc} summarizes the conclusions and outlines potential directions for future research.

\section{Datasets and Methodology}\label{sec_data_meth}
In this section, we describe the observational datasets used this analysis. We also explain the statistical approach based on Bayesian inference and Markov Chain Monte Carlo (MCMC) techniques, which we use to derive constraints on cosmological parameters. 

\subsection{Strong Gravitational Lensing- Distance Ratio}
The Singular Isothermal Sphere (SIS) model remains a standard framework for studying galaxy-galaxy lensing due to its simplicity and physical relevance. In this model, the lensing galaxy is assumed to have a spherically symmetric mass distribution with a density profile that decreases as the inverse square of the radial distance, $\rho(r) \propto r^{-2}$. The spherical symmetry leads to the formation of two images of the background source. The Einstein radius in this model is expressed as \cite{1992ARA&A..30..311B,1996astro.ph..6001N}
\begin{equation}
    \theta_E = 4\pi \dfrac{\sigma^2_{v,\mathrm{SIS}}}{c^2} \frac{d_{A}^{^{\mathrm{ls}}}}{d_{A}^{^{\mathrm{os}}}},
\end{equation}
where $\sigma_{v,\mathrm{SIS}}$ represents the characteristic velocity dispersion of the lens galaxy. This parameter can be estimated through spectroscopic observations that measure the velocity dispersion within an aperture, denoted by $\sigma_\text{ap}$. Since the observed dispersion depends on the aperture size, it requires rescaling to approximate the central velocity dispersion, $\sigma_{v,e/2}$, at the galaxy’s half-light radius. The aperture $\theta_\text{ap}$, originally measured using a rectangular slit, is converted into an equivalent circular aperture using the relation \cite{1995MNRAS.276.1341J}
\begin{equation}
    \theta_\text{ap} \simeq 1.025 \times \sqrt{\frac{\theta_x \theta_y}{\pi}},
\end{equation}
where $\theta_x$ and $\theta_y$ are the slit dimensions. For uniformity across different observational setups, velocity dispersions are normalized to correspond to half the effective radius $\theta_\text{eff}/2$ using the aperture correction formula \cite{1995MNRAS.276.1341J}:
\begin{equation}
   \sigma_{v,\mathrm{SIS}} = \sigma_{e/2} = \sigma_\text{ap} \left(\frac{\theta_\text{eff}}{2\theta_\text{ap}}\right)^\eta.
\end{equation}
Here, we consider the value of the correction parameter $\eta=-0.066 \pm 0.035$ which is derived by Cappellari et al. \cite{2006MNRAS.366.1126C}. This correction ensures that the velocity dispersion matches the scale of the Einstein radius and improves the accuracy of cosmological analyses using the SIS model.

In this analysis, we adopt a sample of galaxy-scale strong gravitational lensing systems originally compiled by Cao et al. (2015) \cite{2015ApJ...806..185C} and subsequently refined by Chen et al. (2019) \cite{2019MNRAS.488.3745C}. The sample consists of 161 lensing galaxies, primarily classified as early-type (E/S0 morphologies), carefully chosen to exclude systems with substantial substructures or nearby companions that could bias lensing measurements. This dataset integrates multiple surveys: 5 systems from the LSD survey \cite{2002ApJ...568L...5K,2003ApJ...583..606K,2002ApJ...575...87T,2004ApJ...611..739T}, 26 from the Strong Lenses in the Legacy Survey (SL2S) \cite{2011ApJ...727...96R,2013ApJ...777...97S,2013ApJ...777...98S,2015ApJ...800...94S}, 57 from the Sloan Lens ACS Survey (SLACS) \cite{2008ApJ...682..964B,2009ApJ...705.1099A,2010ApJ...724..511A}, 38 from the ``SLACS for the Masses'' extension (S4TM) \cite{2015ApJ...803...71S,2017ApJ...851...48S}, 21 from the BOSS Emission-Line Lens Survey (BELLS) \cite{2012ApJ...744...41B}, and 14 from the BELLS for GALaxy-Ly EmitteR sYstems survey (BELLS-GALLERY) \cite{2016ApJ...824...86S,2016ApJ...833..264S}. 

The sample provides detailed measurements necessary for distance ratio analyses, including the lens redshift ($z_l$), source redshift ($z_s$), Einstein radius ($\theta_E$), aperture velocity dispersion ($\sigma_{\rm ap}$) measured within an angular aperture $\theta_{\rm ap}$, and the half-light angular radius of the lens galaxy ($\theta_{\rm eff}$). The lenses span a redshift range of $0.0624 \leq z_l \leq 1.004$, while the source redshifts range from $0.197 \leq z_s \leq 3.595$. The angular Einstein radius, which depends on the distance ratio $d_\mathrm{A}^{^\mathrm{ls}}/d_\mathrm{A}^{^\mathrm{os}}$ between lens and source and observer and source, provides a robust geometric probe for testing cosmological models, provided an accurate lens model is adopted.

% In this work, we employ the simplest and widely used Singular Isothermal Sphere (SIS) model to interpret the lensing observables. Although more sophisticated models such as the extended power-law profile offer the flexibility to describe detailed mass distributions and their redshift evolution (Koopmans 2006; Chen et al. 2019), we choose the SIS model as it sufficiently captures the large-scale geometry required to constrain cosmic curvature and cosmographic parameters. Our objective is to utilize the distance ratio as a cosmological probe without introducing uncertainties from complex lens modeling, thereby focusing solely on the geometric aspects of the universe’s expansion.

\subsection{Late-Time Distance Probes: BAO and SNIa}

In this analysis, we incorporate late-time distance probes from Baryon Acoustic Oscillations (BAO) and Type Ia supernovae (SNIa), which complement the strong lensing distance ratio measurements in constraining cosmological parameters. For a comprehensive description of these datasets, their construction, and associated systematics, we refer the reader to Paper I of this series.

\begin{itemize}

    \item \textbf{Baryon Acoustic Oscillations (BAO):} \\
    We utilize the second-year BAO dataset released by the Dark Energy Spectroscopic Instrument (DESI), corresponding to three years of observations \cite{2025arXiv250314743D,2025arXiv250314738D}. The dataset includes the full covariance matrix to account for correlations among distance indicators. The BAO observables are expressed as distance ratios normalized by the sound horizon at the drag epoch, namely $d_\mathrm{co}(z)/r_d$, $d_H(z)/r_d$, and $d_V(z)/r_d$, which are sensitive to the cosmic expansion history and early universe physics  \cite{2010dken.book..246B,1998ApJ...496..605E,1972CoASP...4..173S}. In the BAO analysis, a strong degeneracy appears between the Hubble constant $H_0$ and the sound horizon $r_d$. To break this degeneracy, we apply a Gaussian prior on $r_d$ using the values reported in reference \cite{2025JCAP...02..021A}.
    \item \textbf{Type Ia Supernovae (SNIa):} \\
    For the supernova observations, we analyze three independent compilations: PantheonPlus \cite{2022ApJ...938..113S}, Union3 \cite{2025ApJ...986..231R}, and DESY5 \cite{2024ApJ...973L..14D}. The PantheonPlus dataset consists of $1701$ light curves from $1550$ spectroscopically confirmed SNIa spanning the redshift range $0.001 \leq z \leq 2.26$ \cite{2022ApJ...938..113S}. To minimize the effects of peculiar velocities on the Hubble diagram, we exclude samples with $z\leq0.01$ and use 1590 light curves over $0.01 < z < 2.26$, which have been carefully corrected for observational biases and cross-calibrated across multiple surveys \cite{2019ApJ...874..150B,2021ApJ...909...26B,2021ApJ...913...49P,2023ApJ...945...84P}. Union3 provides 2087 supernovae, using Bayesian hierarchical modelling to address systematic uncertainties \cite{2025ApJ...986..231R}. At present, only the binned distance modulus measurements are publicly available for this sample; accordingly, we restrict our analysis to these data points and incorporate their corresponding covariance matrix. The DESY5 sample, based on the full five-year Dark Energy Survey, contains $1635$ photometrically classified supernovae in the redshift range $0.0596 < z < 1.12$, along with $194$ low-redshift supernovae from more recent surveys \cite{2024ApJ...973L..14D}. These datasets are treated separately in our analysis to avoid cross-correlations and to test the robustness of cosmological parameter estimates. 
\end{itemize}

\subsection{Methodology}

\begin{itemize}

    \item \textbf{Cosmographic Framework} \\
    The cosmographic approach extracts information directly from observations by assuming only homogeneity and isotropy, without reference to a specific cosmological model. The metric is described by the FLRW line element:
    \begin{equation}
        ds^2 = -c^2 dt^2 + a^2(t)\left[\frac{dr^2}{1-k r^2} + r^2 d\Omega^2 \right],
    \end{equation}
    where $c$ is the speed of light, $k$ represents spatial curvature, and $a(t)$ is the scale factor normalized to unity at the present time. The scale factor is expanded as a Taylor series around the present epoch:
    \begin{equation}
        \frac{a(t)}{a(t_0)} = 1 + H_0(t - t_0) - \frac{q_0}{2} H_0^2 (t - t_0)^2 + \frac{j_0}{3!} H_0^3 (t - t_0)^3 + \frac{s_0}{4!} H_0^4 (t - t_0)^4 + \dots
    \end{equation}
    where the cosmographic parameters are defined as
    \begin{equation}
        H_0 = \left. \frac{\dot{a}}{a} \right|_{t=t_0}, \quad q_0 = -\left. \frac{1}{H_0^2}\frac{\ddot{a}}{a} \right|_{t=t_0}, \quad j_0 = \left. \frac{1}{H_0^3}\frac{\dot{\ddot{a}}}{a} \right|_{t=t_0}, \quad s_0 = \left. \frac{1}{H_0^4}\frac{\ddot{\ddot{a}}}{a} \right|_{t=t_0}.
    \end{equation}
    To ensure convergence across a wide redshift range, we expand the Hubble parameter in terms of the variable $y = \frac{z}{1+z}$ as
    \begin{equation}
        H(y) = H_0 + \left.\frac{dH}{dy}\right|_{y=0}y + \frac{1}{2}\left.\frac{d^2H}{dy^2}\right|_{y=0}y^2 + \frac{1}{6}\left.\frac{d^3H}{dy^3}\right|_{y=0}y^3 + \dots
    \end{equation}
    The distances such as angular diameter distance $d_A$ and comoving distance $d_\mathrm{co}$ can be expressed as functions of these cosmographic parameters. For details on the expansions and their application, we refer the reader to Paper I.

    \item \textbf{Distance Sum Rule (DSR)} \\
    The DSR provides a geometric relation among the dimensionless comoving distances between the observer, lens, and source, denoted as
    \begin{equation}
    \begin{aligned}
        D_\mathrm{co}^{^\mathrm{os}} &= \frac{H_0}{c} d_\mathrm{co}^{^\mathrm{os}}, \\
        D_\mathrm{co}^{^\mathrm{ol}} &= \frac{H_0}{c} d_\mathrm{co}^{^\mathrm{ol}}, \\
        D_\mathrm{co}^{^\mathrm{ls}} &= \frac{H_0}{c} d_\mathrm{co}^{^\mathrm{ls}}.
    \end{aligned}
    \end{equation}
    In a flat universe, these distances satisfy $D_\mathrm{co}^{^\mathrm{os}} = D_\mathrm{co}^{^\mathrm{ol}} + D_\mathrm{co}^{^\mathrm{ls}}$. For a non-flat universe, the relation generalizes to:
    \begin{equation}\label{equ_dsr}
        D_\mathrm{co}^{^{\mathrm{ls}}} = D_\mathrm{co}^{^{\mathrm{os}}} \sqrt{1 + \Omega_{k0} \, \left(D_\mathrm{co}^{^{\mathrm{ol}}}\right)^2} - D_\mathrm{co}^{^{\mathrm{ol}}} \sqrt{1 + \Omega_{k0} \, \left(D_\mathrm{co}^{^{\mathrm{os}}}\right)^2}
    \end{equation}
Using the expression of DSR given in Equation \ref{equ_dsr}, one can rewrite it in terms of distance ratio as
    
    \begin{align}\label{equ_dsr_dr}
        d_\mathrm{AR}^{\mathrm{th}} &= \frac{d_A^{^{\mathrm{ls}}}}{d_A^{^{\mathrm{os}}}} = \frac{d_\mathrm{co}^{^{\mathrm{ls}}}}{d_\mathrm{co}^{^{\mathrm{os}}}} = \frac{D_\mathrm{co}^{^{\mathrm{ls}}}}{D_\mathrm{co}^{^{\mathrm{os}}}}= \sqrt{1 + \Omega_{k0} \left(D_\mathrm{co}^{^{\mathrm{ol}}}\right)^{2}} - \frac{D_\mathrm{co}^{^{\mathrm{ol}}}}{D_\mathrm{co}^{^{\mathrm{os}}}} \sqrt{1 + \Omega_{k0} \left(D_\mathrm{co}^{^{\mathrm{os}}}\right)^{2}}
    \end{align}

    This formulation allows direct determination of the curvature parameter $\Omega_{k0}$ from observations or cosmography expressions of $D_\mathrm{co}^{^\mathrm{ol}}$ and $D_\mathrm{co}^{^\mathrm{os}}$ without assuming a fiducial cosmological model. {In this paper, we apply the DSR to distance ratio measurements, where the comoving distances at the lens and source redshifts are estimated directly from the cosmographic series expansion. This choice removes the need for external data interpolation or secondary distance indicators.} This approach complements the analysis of time-delay distances presented in Paper I.   
    
\end{itemize}

\subsection*{Chi-Square Formalism}

\begin{itemize}

    \item \textbf{SGL Distance Ratio:}
    \begin{equation}
        \chi^2_{\mathrm{SGL}}\left(\Omega_{k0},q_0,j_0,s_0\right) = \sum_i \left[ \frac{d_{AR}^{\mathrm{th}} - d_{AR}^{\mathrm{obs}}}{\sigma_{d_{AR}^{\mathrm{obs}}}} \right]^2
    \end{equation}
    where $d_{AR}$ represents the observed and theoretical distance ratios derived from the lensing systems.

Following Chen et al.~\cite{2019MNRAS.488.3745C}, the total uncertainty in the velocity dispersion is given by
\begin{equation}
\left(\Delta \sigma_\mathrm{SIS}^{\mathrm{tot}}\right)^{2} = \left(\Delta \sigma_{e/2}^{\mathrm{stat}}\right)^{2} + \left(\Delta \sigma_{e/2}^{\mathrm{AC}}\right)^{2} + \left(\Delta \sigma_{e/2}^{\mathrm{sys}}\right)^{2},
\end{equation}
where $\Delta \sigma_{e/2}^{\mathrm{stat}}$ denotes the statistical uncertainty from measurements, $\Delta \sigma_{e/2}^{\mathrm{AC}}$ represents the propagated uncertainty from the aperture correction to the half effective radius, and $\Delta \sigma_{e/2}^{\mathrm{sys}}$ accounts for systematic uncertainties.

The aperture correction uncertainty is computed as
\begin{equation}
\left(\Delta \sigma_{e/2}^{\mathrm{AC}}\right)^{2} = \left(\Delta \sigma_{\mathrm{ap}}\right)^{2} \left( \frac{\theta_\mathrm{eff}}{2\theta_\mathrm{ap}} \right)^{2\eta} + \sigma_{\mathrm{ap}}^{2} \left( \frac{\theta_\mathrm{eff}}{2\theta_\mathrm{ap}} \right)^{2\eta} \ln^{2}\left( \frac{\theta_\mathrm{eff}}{2\theta_\mathrm{ap}} \right) \left(\Delta \eta \right)^{2},
\end{equation}
where $\theta_\mathrm{eff}$ and $\theta_\mathrm{ap}$ are the angular half-light and aperture radii respectively, and $\eta$ is the aperture correction parameter.

For the systematic uncertainty, $\Delta \sigma_{e/2}^{\mathrm{sys}}$ reflects potential deviations arising from assumptions on the consistency between dynamical mass and Einstein mass. We adopt a $3\%$ systematic uncertainty on the model-predicted velocity dispersion to account for line-of-sight matter effects, as suggested by Jiang \& Kochanek~\cite{2007ApJ...671.1568J}.   

% Finally, using the total uncertainty in Equation~(1), the relative uncertainty $\delta \sigma_\mathrm{ap}$ in Equation~(10) is calculated by
% \begin{equation}
% \delta \sigma_\mathrm{ap} = \frac{\Delta \sigma_{e/2}^{\mathrm{tot}}}{\sigma_{e/2}}.
% \end{equation}

    \item \textbf{SNIa:} \\
    The theoretical apparent magnitude for SNIa is given by:

\begin{align}\label{equ_m_th}
    m^{\mathrm{th}} = 5\log_{10}\left(D_L\right) + \mathcal{M},
\end{align}

where $\mathcal{M}=5\log_{10}\left(\dfrac{c}{H_0}\right)+M_B+25$. %In this analysis, we treat $\mathcal{M}$ as a nuisance parameter and numerically marginalize over it using a flat, uniform prior distribution spanning from $-\infty$ to $+\infty$. \\
{To remove dependence on the calibration nuisance parameter $\mathcal{M}$, we perform an analytical marginalization assuming a uniform prior distribution spanning from $-\infty$ to $+\infty$. This exact integration eliminates the need for a specific numerical grid search or an arbitrary prior range truncation.} \\  

After marginalizing over $\mathcal{M}$, the chi-square for SNIa data is:
    \begin{equation}
        \chi^2_{\mathrm{SNIa}}\left(\Omega_{k0},q_0,j_0,s_0\right) = a - \frac{b^2}{f} + \ln\left(\frac{f}{2\pi}\right),
    \end{equation}
    % where
    % \begin{align}
    %     a &= \Delta\bar{m}^T \mathrm{Cov}^{-1} \Delta\bar{m}, \nonumber\\
    %     b &= \Delta\bar{m}^T \mathrm{Cov}^{-1} I, \nonumber\\
    %     f &= I^T \mathrm{Cov}^{-1} I, \nonumber\\
    %     \Delta\bar{m} &= m^{\mathrm{obs}} - 5\log_{10}(D_L).
    % \end{align}
    % For further details regarding the formulation and treatment of systematic uncertainties, see Paper I.
    {where $a$, $b$, and $f$ are defined as $a = \Delta\bar{m}^T \mathrm{Cov}^{-1} \Delta\bar{m}$, $b = \Delta\bar{m}^T \mathrm{Cov}^{-1} I$, and $f = I^T \mathrm{Cov}^{-1} I$. Here, $\Delta\bar{m} = m^{\mathrm{obs}} - 5\log_{10}(D_L)$, and $I$ denotes a vector of ones (or the identity matrix in the covariance formalism). This form incorporates both statistical and systematic uncertainties. For a detailed derivation and discussion of the marginalization procedure, we refer the reader to Paper I.}

    \item \textbf{DESI-DR2:}
    The Chi-square expression for this sample is given as
    \begin{equation}
        \chi^2_{\mathrm{DESI}}\left(H_0,\Omega_{k0},q_0,j_0,s_0\right) = \left[X_{\mathrm{obs}} - X_{\mathrm{th}}\right]^T \mathrm{Cov}^{-1} \left[X_{\mathrm{obs}} - X_{\mathrm{th}}\right],
    \end{equation}
    where $X$ corresponds to the set of distance observables normalized by the sound horizon.

    \item \textbf{Joint Analysis:}
    \begin{equation}
        \chi^2_{\mathrm{Joint}} = \chi^2_{\mathrm{SGL}} + \chi^2_{\mathrm{SNIa}} + \chi^2_{\mathrm{DESI}},
    \end{equation}
    where $\chi^2_{\mathrm{SNIa}}$ applies separately to the PantheonPlus, Union3, and DESY5 datasets.

\end{itemize}

For the cosmological parameters considered in this analysis, we adopt flat priors within specified ranges, as listed in Table~\ref{tab_prior}. The exploration of the parameter space is performed using the Markov Chain Monte Carlo (MCMC) method implemented through the \textbf{\textit{emcee}} sampler \cite{2013PASP..125..306F}. The sampler is configured with 12 walkers, each running for 10,000 steps to to achieve a thorough exploration of the posterior distribution. The initial 30\% of samples are treated as the burn-in phase and excluded from the final analysis. The convergence of this sampler is monitored by computing the integrated auto-correlation time, $\tau_f$, utilizing the \textbf{\textit{autocorr.integrated\_time}} function provided by the \textbf{\textit{emcee}} package. This procedure helps guarantee that the chains have mixed well and the results are statistically robust.

\begin{table}[h]
    \centering
    \renewcommand{\arraystretch}{2}
    \begin{tabular}[b]{|l|l|}
        \hline
        Parameter & Prior Range \\
        \hline \hline
        $H_0~[\mathrm{km\,s^{-1}\,Mpc^{-1}}]$ & $\mathbb{U}[0,100]$ \\
        \hline
        $\Omega_{k0}$ & $\mathbb{U}[-0.5,1]$ \\
        \hline
        $q_0$ & $\mathbb{U}[-2,2]$ \\
        \hline
        $j_0$ & $\mathbb{U}[-3,3]$ \\
        \hline
        $s_0$ & $\mathbb{U}[-10,10]$ \\
        \hline
    \end{tabular}
    \caption{Flat priors assumed for the Hubble constant, spatial curvature, and cosmographic parameters.}
    \label{tab_prior}
\end{table}

{The prior ranges listed in Table~\ref{tab_prior} cover sufficiently broad intervals based on standard cosmological literature to prevent any boundary bias during the sampling process. To ensure that these boundaries do not artificially shift our Bayesian inference, we tested the sensitivity of our constraints using different prior setups. The full discussion of this prior sensitivity analysis, along with the corresponding numerical comparison table for the narrow prior configuration is detailed in the Appendix \ref{sec:appendix_prior_sensitivity}. %For instance, narrowing the Hubble constant prior to $H_0 \in [40, 80]$ returns parameter estimates that are perfectly consistent with our primary results within the $68\%$ confidence interval. 
}

\section{Results}\label{sec_resu}

In this section, we present the constraints on cosmological and cosmographic parameters obtained from the combination of strong gravitational lensing with Type Ia supernova datasets and DESI-DR2 measurements. We analyse three different supernova samples, namely PantheonPlus, Union3, and DESY5, for cases without the inclusion of DESI-DR2 and with the inclusion of DESI-DR2. The estimated values for the Hubble constant ($H_0$), spatial curvature parameter ($\Omega_{k0}$), deceleration parameter ($q_0$), jerk parameter ($j_0$), and snap parameter ($s_0$) are shown in Table~\ref{tab:cosmography_results}. The joint posterior distributions and correlation matrices are shown in Figures~\ref{fig_dr_combined_contour_corr_panthp}, \ref{fig_dr_combined_contour_corr_union3}, and \ref{fig_dr_combined_contour_corr_desy5}.  

\subsection{Constraints without DESI-DR2}

Table~\ref{tab:cosmography_results_1} presents the constraints from the SGL dataset combined with each supernova sample, excluding DESI-DR2. For the SGL+PantheonPlus combination, the curvature parameter is constrained as $\Omega_{k0} = 0.049^{+0.083}_{-0.077}$ which is consistent with a flat universe within the $68\%$ confidence level. The deceleration parameter $q_0 = -0.472^{+0.073}_{-0.067}$ confirms cosmic acceleration with moderate precision. The jerk parameter $j_0 = 0.774^{+0.705}_{-0.737}$ overlaps with the standard model expectation, whereas the snap parameter $s_0 = -0.749^{+6.995}_{-6.395}$ remains largely unconstrained.\\

For the SGL+Union3 combination, the best fit value of $\Omega_{k0}$ i.e. $0.065^{+0.087}_{-0.082}$ shows a slight preference for an open universe but remains compatible with flat geometry at 68\% confidence level. The deceleration parameter $q_0 = -0.272^{+0.151}_{-0.121}$ shows a weaker acceleration, while the jerk parameter $j_0 = -0.372^{+1.029}_{-1.196}$ shows broad uncertainties. The snap parameter $s_0 = -3.327^{+7.753}_{-4.832}$ is poorly constrained.\\

For the SGL+DESY5 dataset, the curvature is $\Omega_{k0} = -0.031^{+0.086}_{-0.075}$, consistent with flat universe at 68\% confidence level. The deceleration parameter $q_0 = -0.735^{+0.141}_{-0.135}$ suggests stronger acceleration. The jerk parameter $j_0 = 2.248^{+0.857}_{-0.933}$ indicates a possible deviation from unity, although the 95\% confidence level interval includes unity. The snap parameter $s_0 = -1.113^{+7.250}_{-6.240}$ remains unconstrained.

\begin{table}[htbp]
\renewcommand{\arraystretch}{2}
\centering
\begin{tabular}{|l|c|c|c|c|}
\hline    
Dataset &  $\Omega_{k0}$ & $q_0$ & $j_0$ & $s_0$ \\
\hline
SGL+PantheonPlus & $0.049^{+0.083}_{-0.077}$ & $-0.472^{+0.073}_{-0.067}$ & $0.774^{+0.705}_{-0.737}$ & $-0.749^{+6.995}_{-6.395}$ \\
\hline
SGL+Union3 & $0.065^{+0.087}_{-0.082}$ & $-0.272^{+0.151}_{-0.121}$ & $-0.372^{+1.029}_{-1.196}$ & $-3.327^{+7.753}_{-4.832}$ \\
\hline
SGL+DESY5 & $-0.031^{+0.086}_{-0.075}$ & $-0.735^{+0.141}_{-0.135}$ & $2.248^{+0.857}_{-0.933}$ & $-1.113^{+7.250}_{-6.240}$ \\
\hline
\end{tabular}
\caption{Parameter constraints from SGL combined with different supernova datasets.}
\label{tab:cosmography_results_1}
\end{table}

\subsection{Constraints with DESI-DR2}

The inclusion of DESI-DR2 considerably tightens the parameter constraints for all dataset combinations, as shown in Table~\ref{tab:cosmography_results}. For SGL+PantheonPlus+DESI-DR2, the Hubble constant is determined as $H_0 = 67.859^{+0.469}_{-0.438}$ km\,s$^{-1}$\,Mpc$^{-1}$, closely aligned with results from other cosmological probes. The curvature parameter tightens to $\Omega_{k0} = 0.139^{+0.061}_{-0.058}$ that is consistent with flat universe at 95\% confidence level. The deceleration parameter $q_0 = -0.481^{+0.057}_{-0.054}$ confirms cosmic acceleration with improved precision. The jerk parameter $j_0 = 1.163^{+0.403}_{-0.425}$ remains consistent with the expected standard value within 68\% confidence level. At the same time, the snap parameter $s_0 = 2.544^{+2.314}_{-2.110}$ shows a tighter constraint compared to the previous case.\\

For SGL+Union3+DESI-DR2, the curvature is $\Omega_{k0} = 0.132^{+0.066}_{-0.062}$, compatible with a flat universe. The Hubble constant $H_0 = 66.709^{+0.622}_{-0.636}$ km\,s$^{-1}$\,Mpc$^{-1}$ aligns with recent Planck observations. The deceleration parameter $q_0 = -0.324^{+0.085}_{-0.103}$ narrows the allowed range, and the jerk parameter $j_0 = 0.208^{+0.647}_{-0.488}$ is consistent with standard predictions. The snap parameter $s_0 = -1.452^{+2.537}_{-1.214}$ is better constrained compared to the non-DESI case. \\

For SGL+DESY5+DESI-DR2, the best-fit value of the curvature, $\Omega_{k0} = -0.054^{+0.050}_{-0.046}$, indicates a slight closed universe while still supporting a flat geometry within the 95\% confidence level. The deceleration parameter $q_0 = -0.546^{+0.087}_{-0.081}$ agrees with an accelerating universe, and the jerk parameter $j_0 = 1.344^{+0.628}_{-0.616}$ overlaps with the standard model expectation. The snap parameter $s_0 = 3.335^{+3.951}_{-3.191}$ has broad uncertainties but still provides valuable constraints.

\begin{table}[htbp]
\renewcommand{\arraystretch}{2}
\centering
\begin{tabular}{|l|c|c|c|c|c|}
\hline
Dataset & $H_0$ & $\Omega_{k0}$ & $q_0$ & $j_0$ & $s_0$ \\
\hline
{SGL+PantheonPlus+DESI-DR2} & $67.859^{+0.469}_{-0.438}$ & $0.139^{+0.061}_{-0.058}$ & $-0.481^{+0.057}_{-0.054}$ & $1.163^{+0.403}_{-0.425}$ & $2.544^{+2.314}_{-2.110}$ \\
\hline
{SGL+Union3+DESI-DR2} & $66.709^{+0.622}_{-0.636}$ & $0.132^{+0.066}_{-0.062}$ & $-0.324^{+0.085}_{-0.103}$ & $0.208^{+0.647}_{-0.488}$ & $-1.452^{+2.537}_{-1.214}$ \\
\hline
{SGL+DESY5+DESI-DR2} & $68.720^{+0.520}_{-0.633}$ & $-0.054^{+0.050}_{-0.046}$ & $-0.546^{+0.087}_{-0.081}$ & $1.344^{+0.628}_{-0.616}$ & $3.335^{+3.951}_{-3.191}$ \\
\hline
\end{tabular}
\caption{Parameter constraints from SGL combined with different supernova datasets including DESI-DR2.}
\label{tab:cosmography_results}
\end{table}

\subsection{Contour Analysis and Parameter Dependencies}

\begin{figure}[htbp]
    \centering
 
    % -------------------- First Row --------------------
    \subfloat[Joint 68\% and 95\% confidence contours and posterior distributions using SGL+PantheonPlus and SGL+PantheonPlus+DESI-DR2 datasets.]
    {\includegraphics[width=0.9\linewidth]{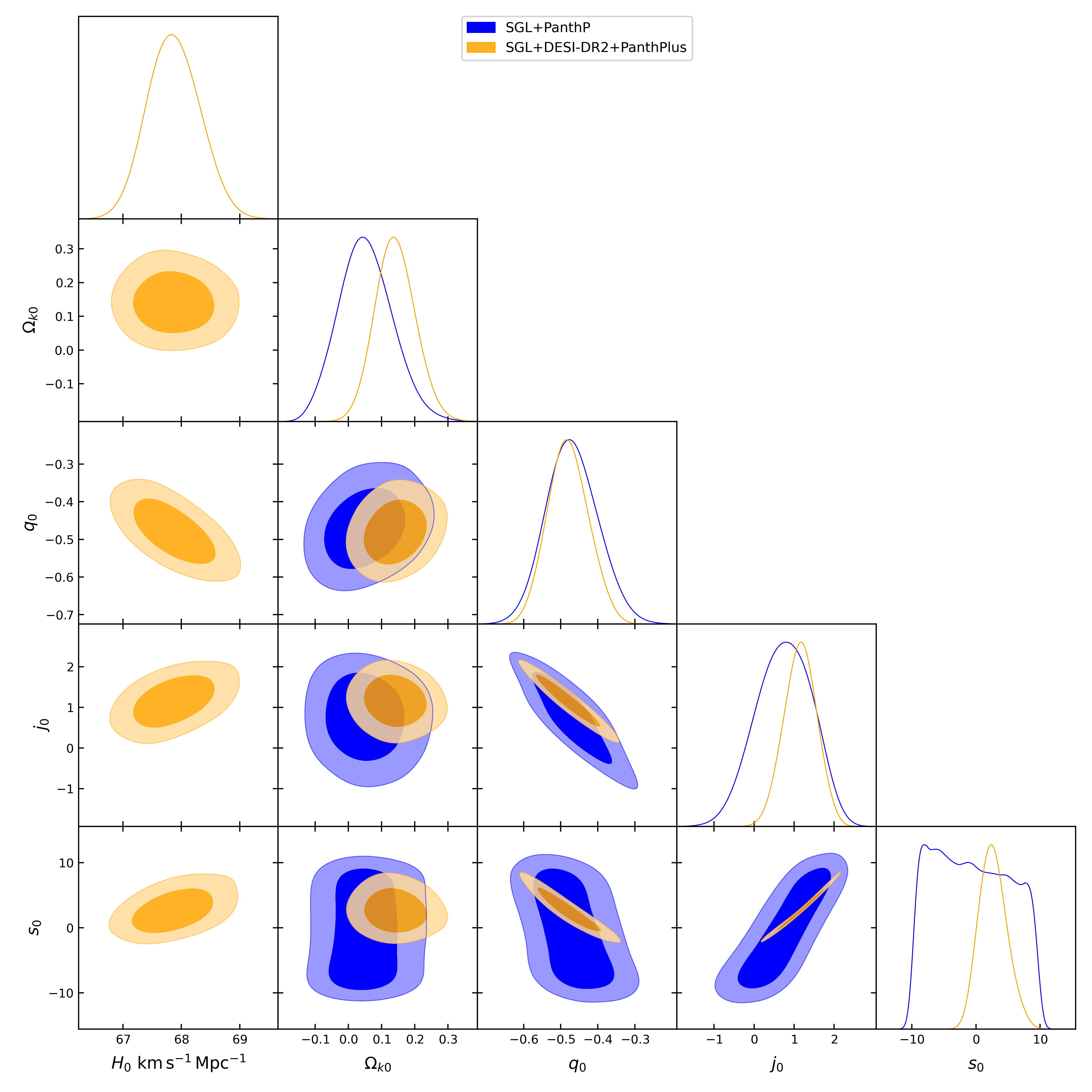}
    \label{fig_dr_cntr_panthp_desi}}\\[6pt]

    % -------------------- Second Row --------------------
    \subfloat[Correlation matrix: SGL+PantheonPlus]
    {\includegraphics[width=0.4\linewidth]{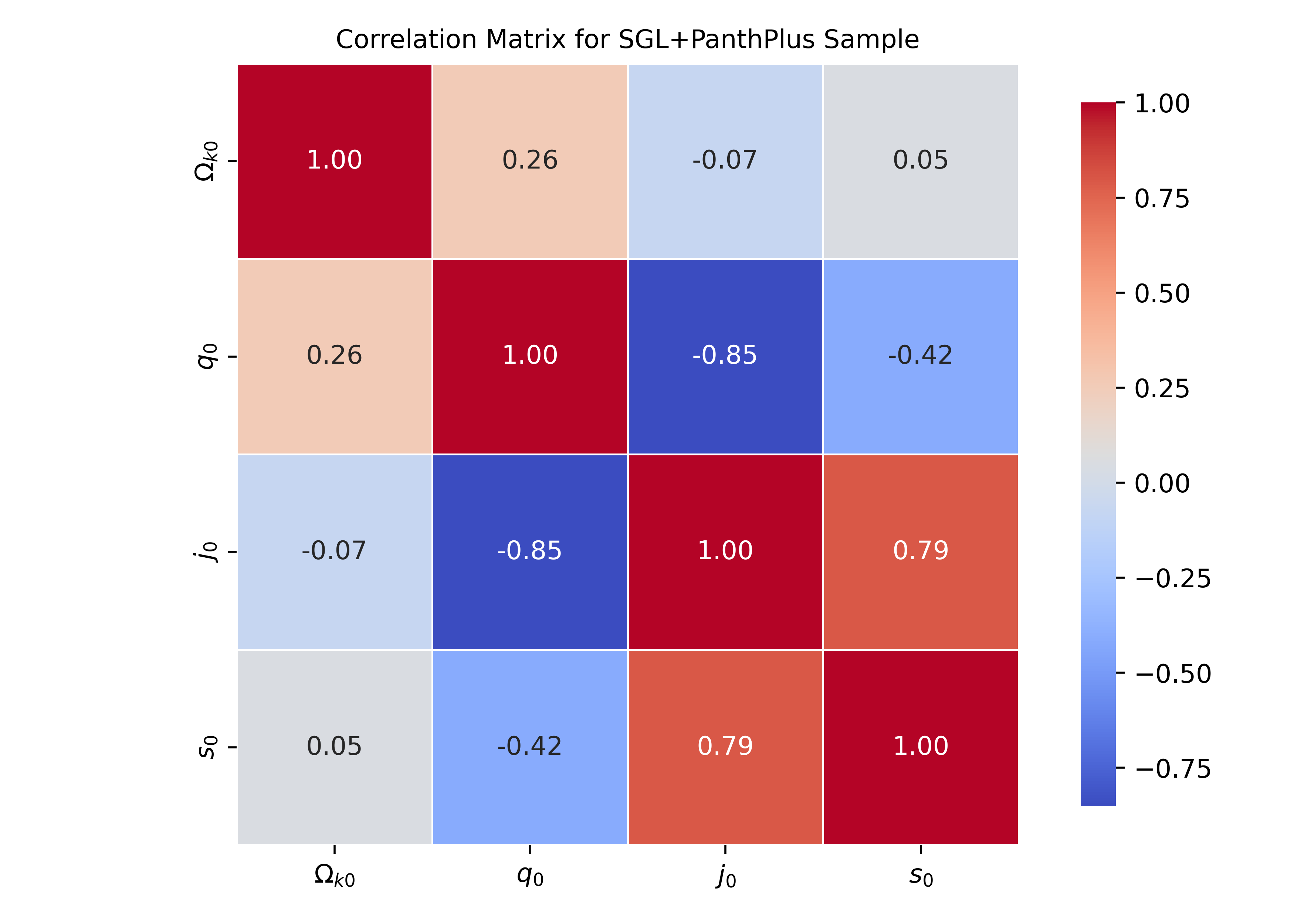}
    \label{fig_dr_corr_panthp}}
    \hspace{0.05\linewidth}
    \subfloat[Correlation matrix: SGL+ PantheonPlus+DESI-DR2]
    {\includegraphics[width=0.4\linewidth]{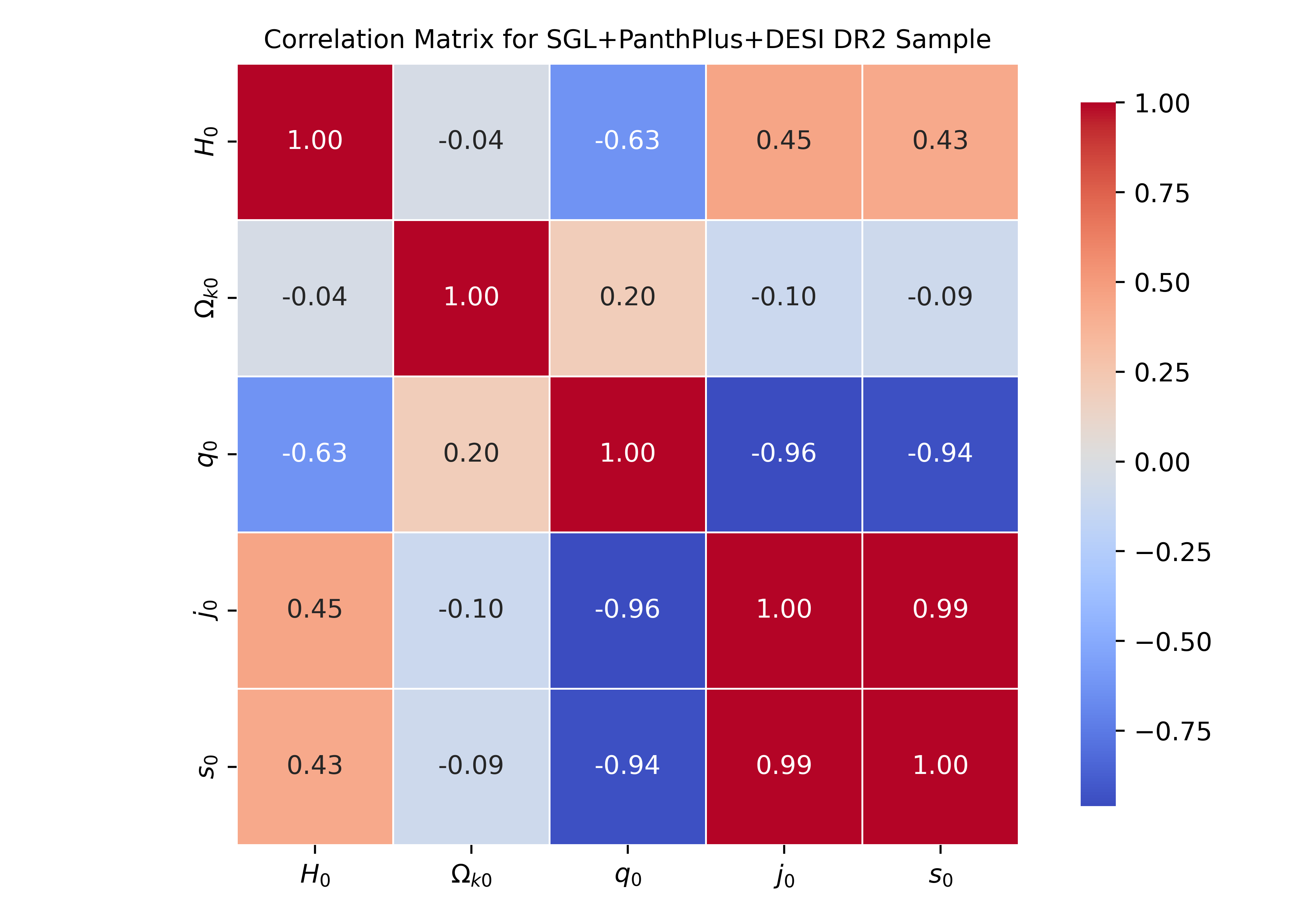}
    \label{fig_dr_corr_panthp_desi}}

    % -------------------- Caption --------------------
       \caption{(a) Posterior contours for SGL+PentheonPlus and SGL+PentheonPlus+DESI-DR2 datasets; (b)–(c) corresponding correlation matrices for each dataset combination.}

    \label{fig_dr_combined_contour_corr_panthp}
\end{figure}

\begin{figure}[htbp]
    \centering

    % -------------------- First Row --------------------
    \subfloat[Joint 68\% and 95\% confidence contours and posterior distributions using SGL+Union3 and SGL+Union3+DESI-DR2 datasets.]
    {\includegraphics[width=0.9\linewidth]{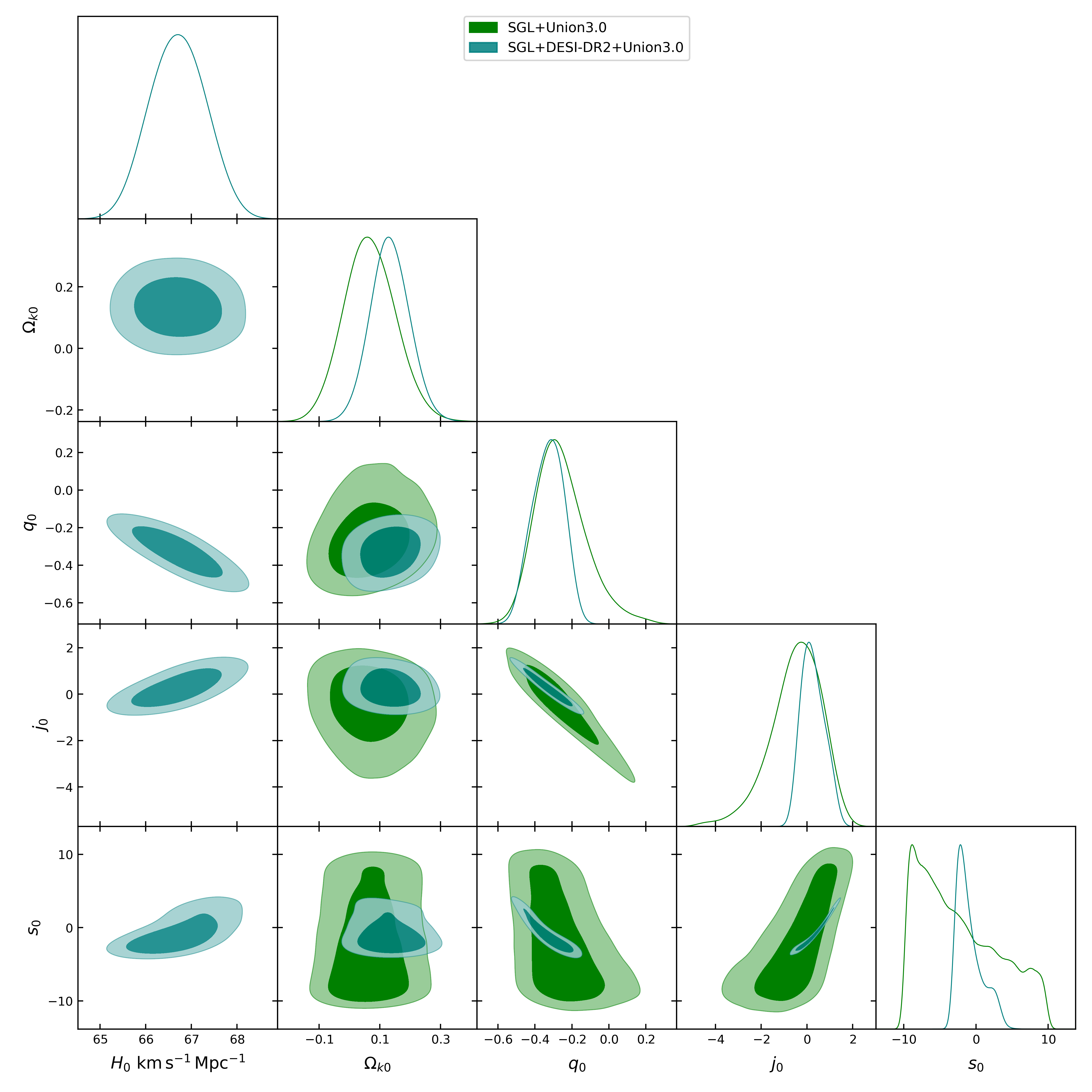}
    \label{fig_dr_cntr_union3_desi}}\\[6pt]

    % -------------------- Second Row --------------------
    \subfloat[Correlation matrix: SGL+Union3]
    {\includegraphics[width=0.4\linewidth]{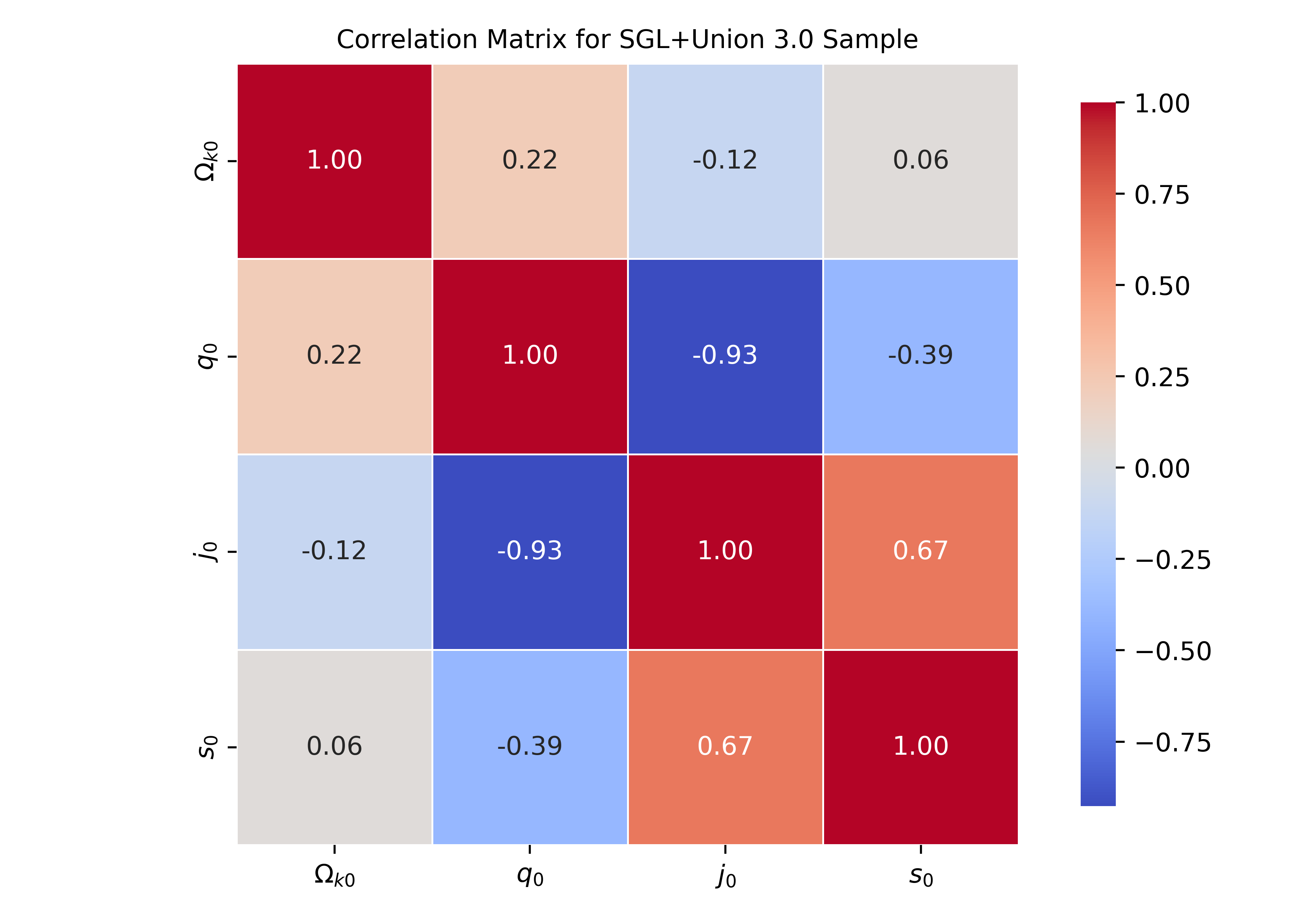}
    \label{fig_dr_corr_union3}}
    \hspace{0.05\linewidth}
    \subfloat[Correlation matrix: SGL+ Union3+DESI-DR2]
    {\includegraphics[width=0.4\linewidth]{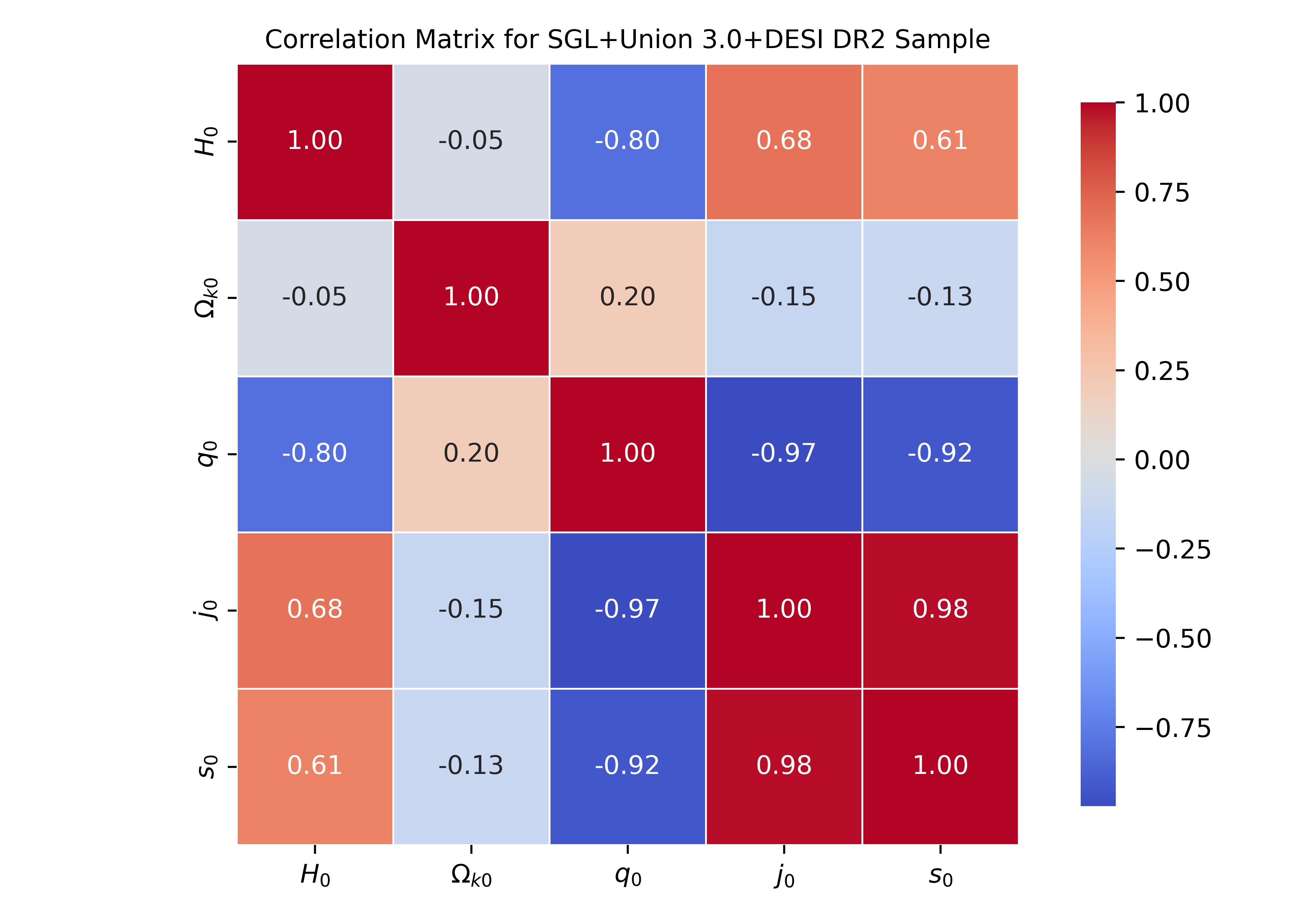}
    \label{fig_dr_corr_union3_desi}}

    % -------------------- Caption --------------------
       \caption{(a) Posterior contours for SGL+Union3 and SGL+Union3+DESI-DR2 datasets; (b)–(c) corresponding correlation matrices for each dataset combination.}

    \label{fig_dr_combined_contour_corr_union3}
\end{figure}

\begin{figure}[htbp]
    \centering

    % -------------------- First Row --------------------
    \subfloat[Joint 68\% and 95\% confidence contours and posterior distributions using SGL+DESY5 and SGL+DESY5+DESI-DR2 datasets.]
    {\includegraphics[width=0.9\linewidth]{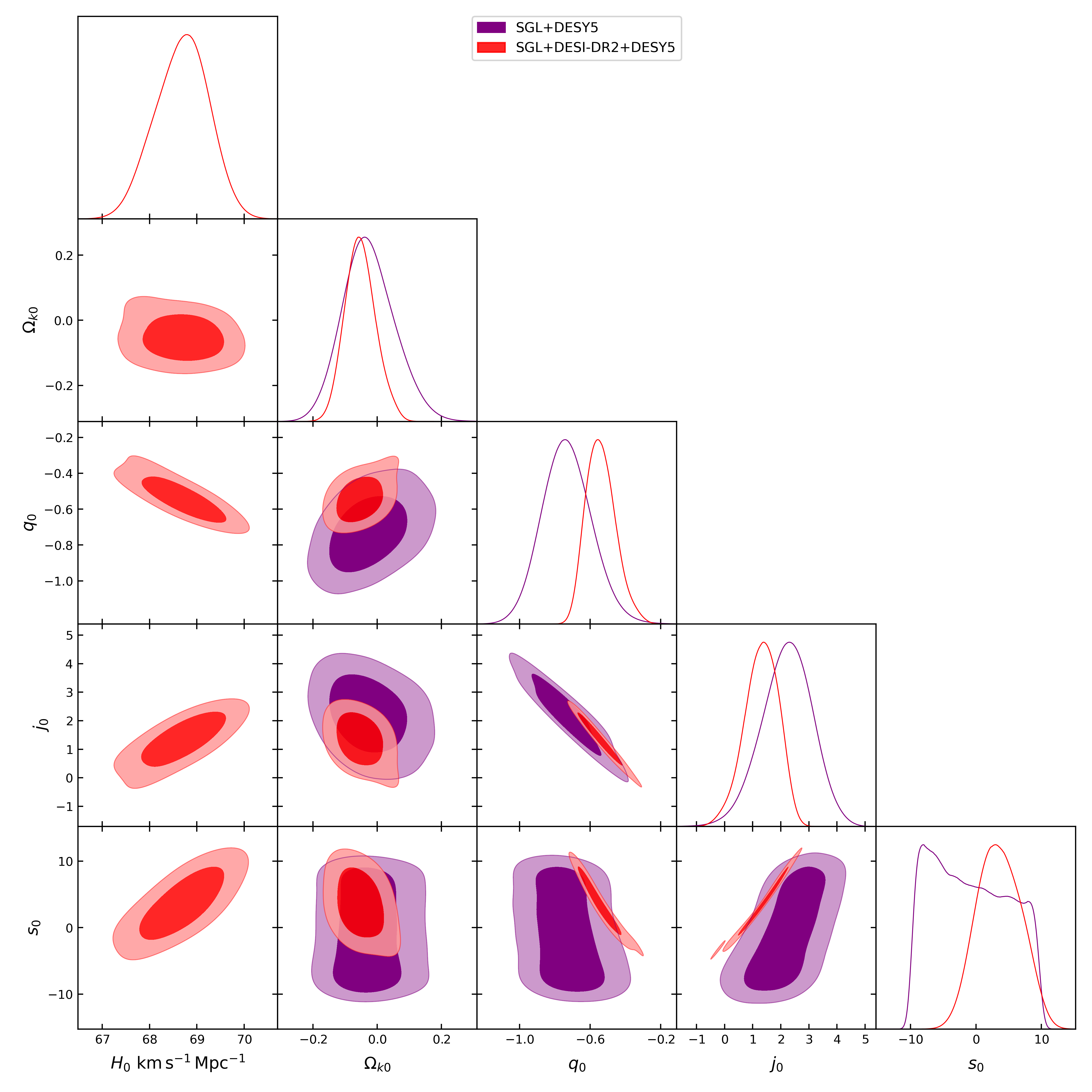}
    \label{fig_dr_cntr_desy5_desi}}\\[6pt]

    % -------------------- Second Row --------------------
    \subfloat[Correlation matrix: SGL+DESY5]
    {\includegraphics[width=0.4\linewidth]{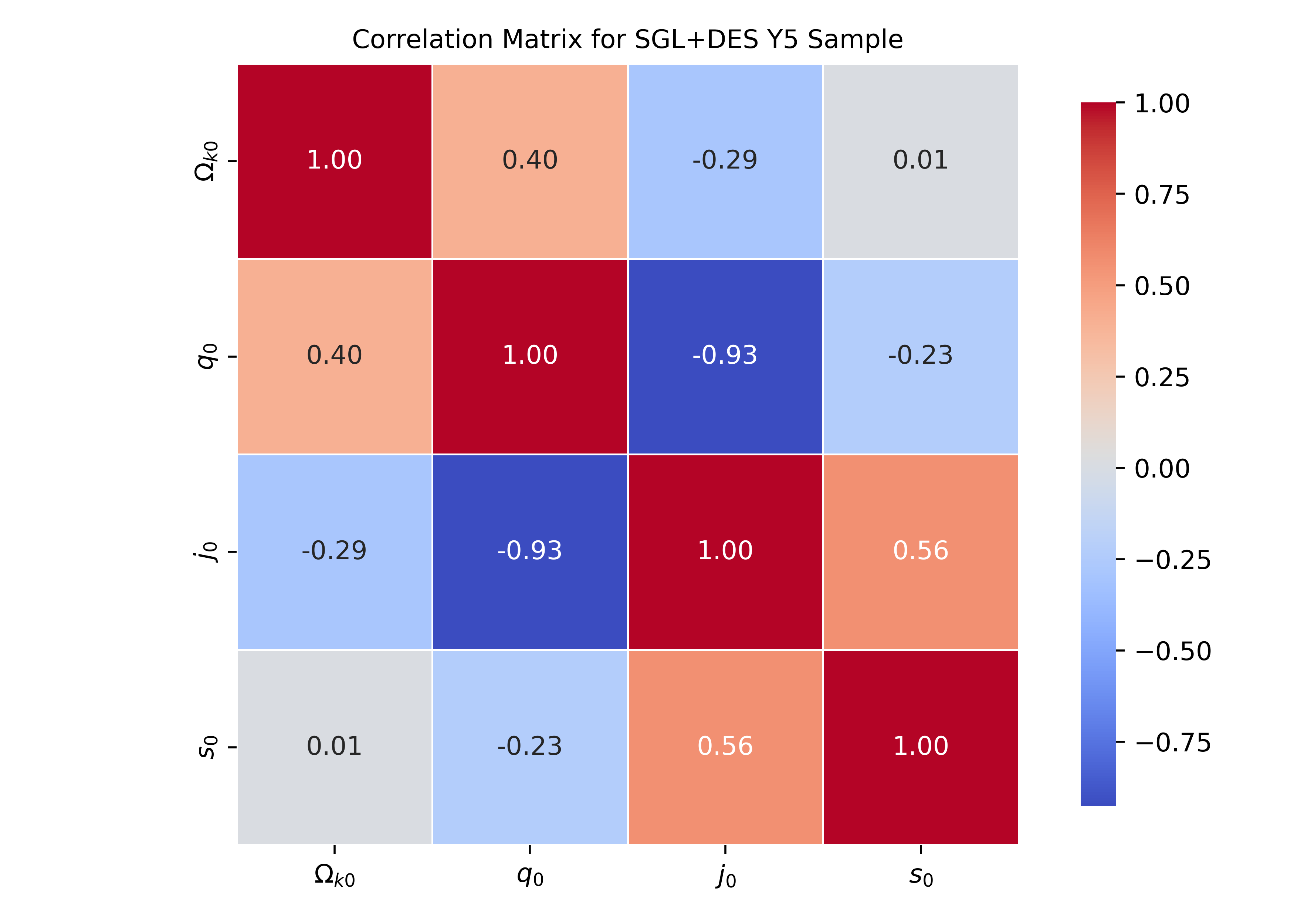}
    \label{fig_dr_corr_desy5}}
    \hspace{0.05\linewidth}
    \subfloat[Correlation matrix: SGL+ DESY5+DESI-DR2]
    {\includegraphics[width=0.4\linewidth]{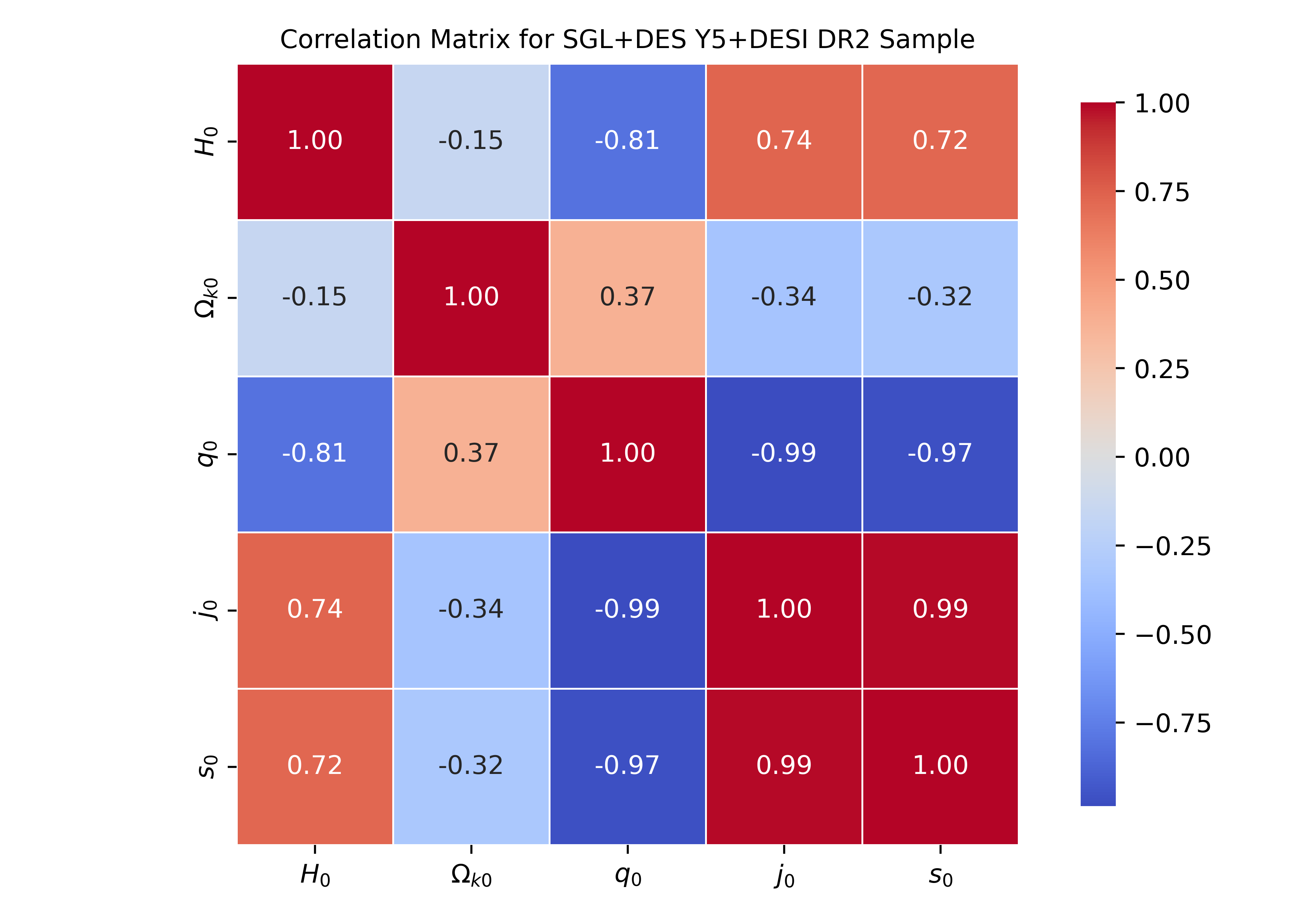}
    \label{fig_dr_corr_desy5_desi}}

    % -------------------- Caption --------------------
       \caption{(a) Posterior contours for SGL+DESY5 and SGL+DESY5+DESI-DR2 datasets; (b)–(c) corresponding correlation matrices for each dataset combination.}

    \label{fig_dr_combined_contour_corr_desy5}
\end{figure}

Figures~\ref{fig_dr_combined_contour_corr_panthp}, \ref{fig_dr_combined_contour_corr_union3}, and \ref{fig_dr_combined_contour_corr_desy5} show the joint posterior distributions with $68\%$ and $95\%$ confidence regions for the considered parameters. These figures clearly demonstrate the impact of DESI-DR2 on tightening the constraints and reducing parameter uncertainities.\\

% \begin{figure}[htbp]
%     \centering
%     \includegraphics[width=0.98\linewidth]{images/DSR_SGL_Csmgrphy_DR_contour_panthp_and_DESI.png}
%     \caption{Joint 68\% and 95\% confidence contours and posterior distributions using SGL+PantheonPlus and SGL+PantheonPlus+DESI-DR2 datasets.}
%     \label{fig_panthp}
% \end{figure}

% \begin{figure}[htbp]
%     \centering
%     \includegraphics[width=0.98\linewidth]{images/DSR_SGL_Csmgrphy_DR_contour_union3_and_DESI.png}
%     \caption{Joint 68\% and 95\% confidence contours and posterior distributions using SGL+Union3 and SGL+Union3+DESI-DR2 datasets.}
%     \label{fig_union3}
% \end{figure}

% \begin{figure}[htbp]
%     \centering
%     \includegraphics[width=0.98\linewidth]{images/DSR_SGL_Csmgrphy_DR_contour_desy5_and_DESI.png}
%     \caption{Joint 68\% and 95\% confidence contours and posterior distributions using SGL+DESY5 and SGL+DESY5+DESI-DR2 datasets.}
%     \label{fig_desy5}
% \end{figure}

Figure~\ref{fig_dr_combined_contour_corr_panthp} shows that, for the PantheonPlus dataset, the constraints improve significantly with DESI-DR2. The contours for $\Omega_{k0}$ and the cosmographic parameters shrink, and this shows a better determination of these parameters. \\ %The deceleration and jerk parameters also benefit from the inclusion of additional distance measurements.\\         

Figure~\ref{fig_dr_combined_contour_corr_union3} presents a similar trend for the Union3 dataset, where the contours without DESI-DR2 exhibit substantial degeneracies between curvature and deceleration. With DESI-DR2, the contours contract and overlap more closely with the expectations from a flat universe.\\

Figure~\ref{fig_dr_combined_contour_corr_desy5} confirms that, even for the DESY5 sample, which shows broader constraints without DESI, the inclusion of DESI-DR2 markedly enhances the precision of all parameters. The tighter contours for $\Omega_{k0}$ and $q_0$ further support the consistency with the $\Lambda$CDM model.\\

\noindent The results show that including DESI-DR2 in cosmographic analyses tightens parameter constraints, improves agreement with standard cosmology, and reduces uncertainties across multiple observational probes.

\section{Discussions and Conclusions}\label{sec_disc_conc}              

In this work, we perform a cosmographic investigation by utilizing strong gravitational lensing distance ratios in conjunction with Type Ia supernova datasets and the DESI-DR2 compilation. Our goal is to refine the estimates of cosmological parameters, including the Hubble constant ($H_0$), spatial curvature parameter ($\Omega_{k0}$), deceleration parameter ($q_0$), jerk parameter ($j_0$), and snap parameter ($s_0$), by performing an analysis that combines diverse observational probes. To put constraints on cosmological parameters, we use Bayesian inference with Markov Chain Monte Carlo (MCMC) methods, marginalize over nuisance parameters, and quantify the confidence regions. Compared with previous studies, this analysis introduces novel ways to assess cosmographic parameters by combining distance ratios from SGL with independent supernova measurements, thereby reducing reliance on traditional distance ladders.

Earlier cosmographic analyses predominantly emphasized distance measurements such as baryon acoustic oscillations or supernova compilations \cite{2010JCAP...03..005V,2020ApJ...900...70R}, while other works incorporated the distance sum rule (DSR) to constrain curvature using heterogeneous datasets \cite{2019MNRAS.483.1104Q,2023MNRAS.521.4963D}. Although some efforts incorporated strong lensing observations, they often depended on calibrated distance measures rooted in local observations \cite{2021PhRvD.103f3511K,2019PhRvL.123w1101C}. \textit{Our approach uniquely combines the distance sum rule with cosmographic expansions and SGL distance ratios, supported by DESI-DR2 data, to construct an independent and robust estimation of the universe’s expansion geometry.} This methodology improves the precision and model-testing capabilities by offering a complementary dataset-based approach that mitigates systematic dependencies.    

{This paper is the second in a two-part study that explores cosmographic constraints from strong gravitational lensing and related datasets. \textbf{Paper I} examined time-delay systems and showed that DESI-DR2 changes the curvature preference: the sign of $\Omega_{k0}$ shifted from positive to negative once DESI-DR2 was included. That analysis also made clear that the observed shift arises from a degeneracy between curvature and late-time expansion. The present paper builds on those results and studies lensing distance ratios, which do not involve $H_0$ as a free parameter. This feature sets them apart from time-delay systems and makes them directly sensitive to curvature and higher-order cosmographic parameters. The combination of these two studies provides complementary tests; time delays set the absolute distance scale, and distance ratios probe the relative geometry, and thus offers a more complete view of late-time cosmology. The primary conclusions of our study are summarized as follows:}

\begin{figure}[htbp]
    \centering
    \includegraphics[width=1\linewidth]{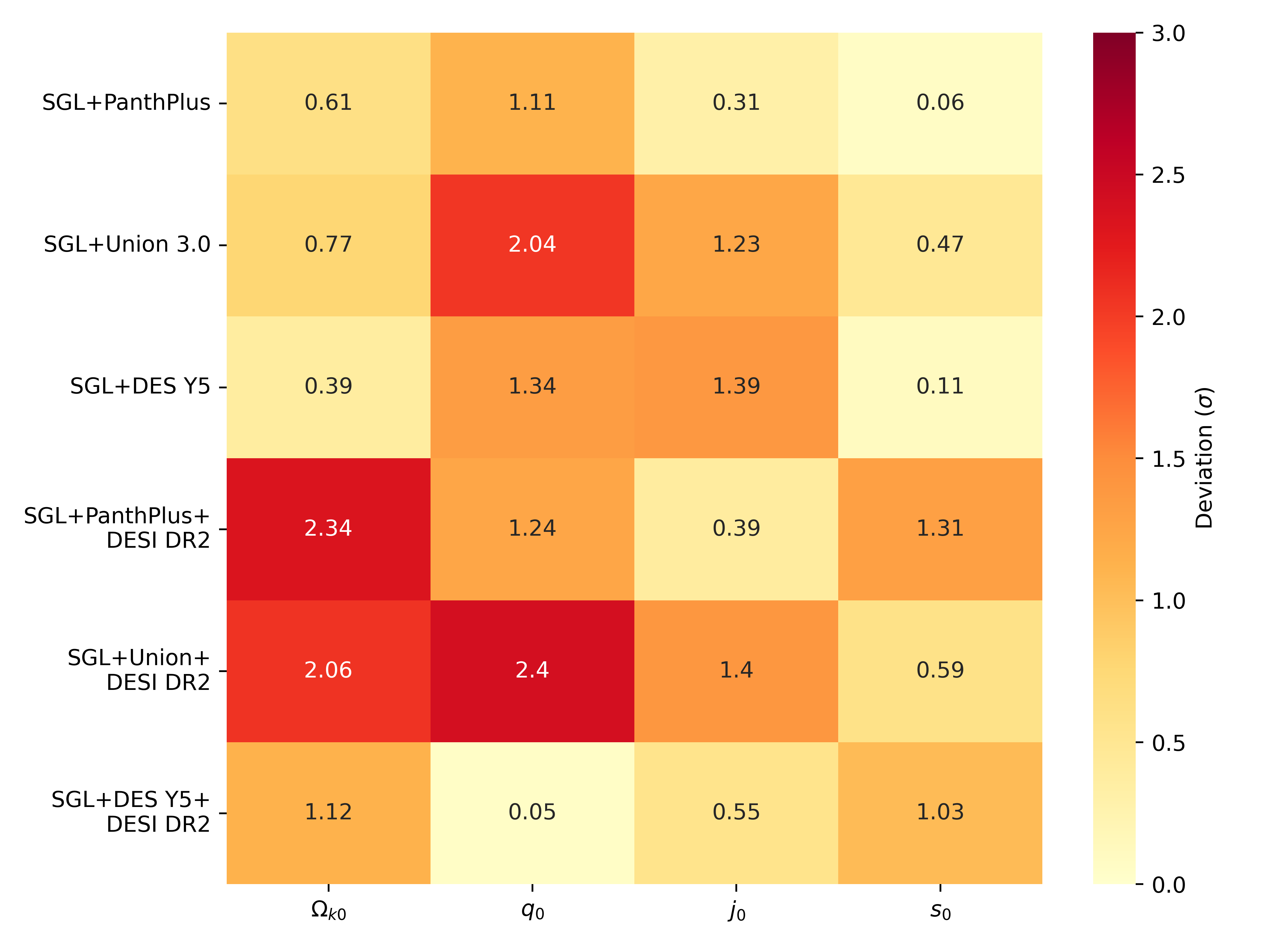}  
    \caption{Tension matrix displaying the deviation from the reference values ($\Omega_{k0}=0$, $q_0 = -0.55$, $j_0 = 1$, $s_0 = -0.35$) in units of $\sigma$ for various observational combinations. The results highlight how the inclusion of DESI-DR2 data systematically reduces deviations, especially for higher-order cosmographic parameters.}       
    \label{fig:sigma_devi}
\end{figure}

\begin{itemize}
    \item \textbf{Analysis without DESI-DR2:} The constraints derived from combining SGL with individual supernova samples such as PantheonPlus, Union3, and DESY5 remain broad, particularly for higher-order cosmographic parameters. The inferred values of $H_0$ tend to exceed those predicted by Planck’s $\Lambda$CDM results ($H_0 \approx 67.4~\mathrm{km\,s^{-1}\,Mpc^{-1}}$), though within obtained error values. The curvature parameter $\Omega_{k0}$ shows hints of deviation but does not significantly contradict flat geometry. The deceleration parameter $q_0$ is less negative, suggesting weaker cosmic acceleration. Uncertainties in $j_0$ and $s_0$ are considerable, with $s_0$ being poorly constrained. The inclusion of distance ratio samples of SGL helps stabilize parameter estimation but cannot fully overcome the statistical limitations posed by the supernova datasets alone.

    \item \textbf{Impact of DESI-DR2:} The integration of DESI-DR2 data substantially improves parameter constraints. The uncertainties shrink by multiple factors, and parameter estimates for $H_0$ align more closely with the standard cosmological model. The curvature parameter is tightly clustered around zero, and the deceleration parameter conforms to expectations of an accelerating universe. The jerk parameter approaches its standard theoretical value, and even the higher-order snap parameter benefits from the additional data, becoming more constrained. The inclusion of DESI-DR2 helps in reduce uncertainties and improves the robustness of the cosmographic results.

    \item \textbf{Comparison with $\Lambda$CDM:} {Without DESI-DR2, the estimates of $q_0$ and $s_0$ show larger uncertainties in $q_0$ and $s_0$ parameters. These differences become smaller when DESI-DR2 data are added. The tension matrix in Figure~\ref{fig:sigma_devi} shows the level of deviation from the reference values for different dataset combinations. The results show that the differences seen without DESI-DR2 mainly come from limited statistical precision and not from any real inconsistency in the cosmological model.}             

    % \item \textbf{Comparison with $\Lambda$CDM:} The absence of DESI-DR2 data leads to larger deviations from $\Lambda$CDM in $q_0$ and $s_0$. These discrepancies reduce once DESI-DR2 is included. The heatmap in Figure~\ref{fig:sigma_devi} illustrates the degree of deviation from reference values across various dataset combinations. The results confirm that discrepancies observed without DESI-DR2 likely arise from insufficient statistical strength rather than fundamental inconsistencies in the cosmological model.

    \item \textbf{Parameter Constraints and Improvements:} The analysis of the joint likelihoods reveals that the parameter space shows large, weakly constrained contours when DESI-DR2 is excluded. This is particularly evident for the parameter $\Omega_{k0}$ and cosmographic parameters, where the extended contours impose fundamental limits on the precision of any resulting constraints. The inclusion of the DESI-DR2 dataset dramatically reduces the area of these contours and produces tightly defined confidence regions. The correlation matrices (Figures \ref{fig_dr_combined_contour_corr_panthp}, \ref{fig_dr_combined_contour_corr_union3}, and \ref{fig_dr_combined_contour_corr_desy5}) further reveal a  modest rise in the correlations among the cosmographic parameters themselves. This pattern signifies that the dataset requires a more coherent and unified constraint across the entire parameter space. This overall refinement is supported by the distance ratios, which improves the stability and precision of the cosmographic inferences. The consistency between the tightened contours and the correlation matrices offers a unified view of the significant improvement achieved by including DESI-DR2.
    
    % \item \textbf{Parameter degeneracies and improvements:} The contours derived from our joint likelihood analysis reveal wide correlations between parameters when DESI-DR2 is excluded, particularly between $H_0$ and $\Omega_{k0}$, and between $q_0$ and $s_0$. The addition of DESI-DR2 data sharpens these contours, resulting in well-defined confidence regions and reducing parameter correlations. The SGL distance ratios contribute significantly by establishing the distance measurements, which improves the stability and precision of the cosmographic inferences.     

\end{itemize}

{ To ensure that our cosmographic results are robust against choices made during the lens modeling stage, we also look into potential systematic shifts from the idealized Singular Isothermal Sphere (SIS) mass profile. Following standard approaches, we test a configuration where a free velocity bias parameter $f_E$ is introduced and marginalized over using a flat prior distribution. We find that the resulting kinematic parameter boundaries are almost entirely insensitive to this structural modification, remaining perfectly consistent with our baseline calculations within the $1\sigma$ standard errors. A full breakdown of this validation test, along with the corresponding numerical comparison table for the SGL+PantheonPlus+DESI-DR2 dataset, is detailed in the Appendix \ref{sec:appendix_lens_profile}. }

While our results offer valuable perspective into cosmographic parameter estimation, it is essential to recognize the limitations inherent in the datasets and methodology. Systematic effects arising from lens mass modeling, environmental contamination, or selection biases in supernova surveys may introduce uncertainties. Furthermore, the significant error bars in parameters such as $j_0$ and $s_0$ suggest that current datasets, although informative, are not sufficient to fully characterize higher-order dynamics in the universe’s expansion. Future analyses should incorporate refined modeling approaches and extended datasets to better isolate these effects and ensure reliability.

For future work, several directions can further enhance the cosmographic framework. Incorporating additional datasets such as baryon acoustic oscillations, cosmic chronometers, and weak lensing observations would provide complementary constraints and reduce degeneracies. It will be critical for future precision cosmology to better understand and mitigate systematic uncertainties, particularly in lens modeling and supernova calibration. It is also possible that the exploration of extended cosmographic formulations or alternative gravity theories will provide a deeper understanding of dark energy and cosmic acceleration. %Efforts to better understand and mitigate systematic uncertainties, particularly in lens modeling and supernova calibration, will be critical for future precision cosmology. Exploration of extended cosmographic formulations or alternative gravity theories may provide deeper understanding of dark energy and cosmic acceleration. 
The application of sophisticated statistical techniques, including machine learning algorithms, may also prove instrumental in handling complex parameter spaces and extracting robust constraints.

In conclusion, our analysis demonstrates that the integration of strong lensing distance ratios with supernova observations, especially when supplemented by DESI-DR2 data, produces tighter and more consistent cosmographic constraints. These results not only support the standard $\Lambda$CDM model but also establish a framework for future explorations into the universe’s expansion history. The methodology developed here sets the way for more accurate and independent assessments of cosmological parameters, thus laying the groundwork for upcoming observational datasets and theoretical advancements.     

\section*{Data Availability}              
\noindent This research did not yield any new data.
      
\section*{Conflict of Interest} 
\noindent The authors declare no conflict of interest.

\section*{Acknowledgments} 
We thank the anonymous referee for helpful comments that improved this work. Darshan Kumar is supported by the Henan Provincial Natural Science Foundation under Grant No. 262300421843 and by the Startup Research Fund of the Henan Academy of Sciences under Grant No. 241841219. One of the authors (Deepak Jain) thanks Inter-University Centre for Astronomy and Astrophysics (IUCAA), Pune (India) for the hospitality provided under the associateship programme where part of the work was done. In this work the figures were created with \textbf{{\texttt{GetDist}}}~\cite{2019arXiv191013970L}, \textbf{ {\texttt{numpy}}}~\cite{numpy}  and \textbf{{\texttt{matplotlib}}}~\cite{matplotlib} Python modules and to estimate parameters we used the publicly available MCMC algorithm  \textbf{ {\texttt{emcee}}} \citep{emcee}. 

\appendix   

\section{Sensitivity Analysis and Choice of Prior Boundaries}
\label{sec:appendix_prior_sensitivity}

In this appendix, we address the potential impact of prior parameter boundaries on our Bayesian inference outcomes. When performing cosmological constraints via Markov Chain Monte Carlo (MCMC) simulations, the chosen ranges for flat priors can act as an implicit regularization. If the data is poorly informative, the resulting posterior distributions can depend heavily on these chosen boundaries. 

The initial parameter ranges listed in Table \ref{tab_prior} were selected to be sufficiently broad and uniform. This setup guarantees that the parameter space exploration is driven purely by the observational data, preventing any artificial truncation of the posterior distributions near the edges. 

To test the stability of our cosmographic parameters and evaluate whether the boundaries influence our final values, we perform a sensitivity analysis by altering the prior on the Hubble constant ($H_0$). We replace our baseline uniform prior, $H_0 \in [0, 100]$, with a narrower yet still uniform range, $H_0 \in [40, 80]$. This tighter boundary covers all realistic low-redshift observational constraints while testing the stability of our high-redshift combinations.

We rerun our analysis for the full joint configuration (SGL + PantheonPlus + DESI-DR2) under this alternative setup. The comparison between the parameter constraints obtained with the wide and narrow priors is shown in Table~\ref{tab:prior_sensitivity_comparison}.

\begin{table}[htbp]   
\renewcommand{\arraystretch}{1.5}
\centering
\begin{tabular}{|l|c|c|c|c|c|}
\hline
Prior Configuration & $H_0$ & $\Omega_{k0}$ & $q_0$ & $j_0$ & $s_0$ \\
\hline
Baseline Wide Prior & $67.859^{+0.469}_{-0.438}$ & $0.139^{+0.061}_{-0.058}$ & $-0.481^{+0.057}_{-0.054}$ & $1.163^{+0.403}_{-0.425}$ & $2.544^{+2.314}_{-2.110}$ \\
\hline
Alternative Narrow Prior & $67.841^{+0.468}_{-0.441}$ & $0.136^{+0.062}_{-0.059}$ & $-0.480^{+0.055}_{-0.057}$ & $1.162^{+0.426}_{-0.426}$ & $2.550^{+2.439}_{-2.122}$ \\
\hline                                                 
\end{tabular}
\caption{Comparison of cosmographic parameter constraints extracted from the joint SGL + PantheonPlus + DESI-DR2 dataset using the baseline wide prior range ($H_0 \in [0,100]$) versus the tighter testing range ($H_0 \in [40,80]$).}
\label{tab:prior_sensitivity_comparison}
\end{table}

The numerical comparisons reveal that our cosmological constraints are almost entirely unaffected by the prior boundaries. The estimated values and statistical errors for the Hubble constant $H_0$, spatial curvature $\Omega_{k0}$, and the deceleration parameter $q_0$ are practically identical between the two runs. The higher-order cosmographic parameters, jerk ($j_0$) and snap ($s_0$), also remain completely stable and well within their $1\sigma$ uncertainties.

This test shows that our combined observational datasets carry significant statistical weight. Because the likelihood profile is well localized inside both ranges, the choice of the prior boundary does not dominate or alter the final results. This confirms the stability of the cosmological conclusions presented in this work.      

\section{Robustness of the SIS assumption for strong lensing data} 
\label{sec:appendix_lens_profile}

In this appendix, we evaluate the sensitivity of our cosmographic parameter constraints to the underlying assumptions regarding the lens galaxy mass profile. As explicitly noted in the Introduction, the primary objective of this work is not to investigate the detailed structural properties or internal mass distributions of lensing galaxies, but rather to constrain the model-independent expansion history of the universe using high-redshift datasets. 

Nevertheless, assuming an idealized Singular Isothermal Sphere (SIS) profile without accounting for complex structural dynamics can potentially introduce systematic shifts in cosmological inference \citep{2003MNRAS.343..639O}. To address this and verify the robustness of our main results, we perform a dedicated validation test by introducing a phenomenological velocity bias parameter, $f_E$. This parameter scales the model characteristics to match the observed central velocity dispersion through the relation:
\begin{equation}
\sigma_{\mathrm{SIS}} = f_E \left( \frac{\sigma_e}{2} \right).
\end{equation}

To ensure a conservative test, we treat $f_E$ as a free nuisance parameter and marginalize over it numerically within our MCMC framework, assigning it a wide, uniform prior distribution:
\begin{equation}
f_E \sim U[0, 2].
\end{equation}

We apply this extended modeling framework to our primary joint dataset combination, SGL + PantheonPlus + DESI-DR2. The resulting parameter constraints are presented in Table~\ref{tab:sis_comparison} alongside our baseline SIS results for direct comparison.

\begin{table}[!h]
\renewcommand{\arraystretch}{1.5}
\centering
\caption{Comparison of cosmographic parameter constraints extracted from the joint SGL + PantheonPlus + DESI-DR2 dataset, contrasting the baseline SIS model against the extended framework marginalized over the lens velocity bias parameter $f_E$.}
\label{tab:sis_comparison}
\begin{tabular}{|l|c|c|}
\hline        
Parameter & Baseline Model (SIS) & SIS Model with bias parameter (With $f_E$ Marginalization) \\ 
\hline
$H_0$ & $67.859^{+0.469}_{-0.438}$ & $67.659^{+0.417}_{-0.448}$ \\ \hline
$\Omega_{k0}$ & $0.139^{+0.061}_{-0.058}$ & $0.235^{+0.068}_{-0.065}$ \\ \hline
$q_0$ & $-0.481^{+0.057}_{-0.054}$ & $-0.449^{+0.048}_{-0.050}$ \\ \hline
$j_0$ & $1.163^{+0.403}_{-0.425}$ & $1.035^{+0.378}_{-0.360}$ \\ \hline
$s_0$ & $2.544^{+2.314}_{-2.110}$ & $1.961^{+2.049}_{-1.714}$ \\ \hline
\end{tabular}
\end{table}

The numerical results demonstrate that our core cosmographic parameters are remarkably stable against the addition of this structural degree of freedom. Every single parameter constraint remains completely consistent within the $1\sigma$ standard deviation boundaries across both model configurations. 

The most visible shift occurs in the spatial curvature parameter, which changes from $\Omega_{k0} = 0.139$ to $\Omega_{k0} = 0.235$. This behavior is physically expected and well-documented in previous strong lensing literature \cite{2017ApJ...834...75X}; the geometric scaling of distance ratios introduces an inherent degeneracy between the spatial curvature of the universe and the mass profile parameters of the lens. Crucially, this shift remains entirely within our statistical error bars and does not modify the overall geometric picture. 
 
Meanwhile, the remaining kinematic parameters show almost no sensitivity to the extra parameter: the Hubble constant $H_0$ shifts by less than one percent, the deceleration parameter $q_0$ remains firmly negative, and the higher-order jerk ($j_0$) and snap ($s_0$) parameters stay fully consistent with standard cosmological backgrounds within their respective uncertainties. 

Consequently, while incorporating $f_E$ adds an extra layer of statistical marginalization, it does not alter any of the primary physical conclusions of this work. This check confirms that our baseline framework is perfectly sufficient and reliable for the precision level required by this cosmographic study.

% We thank the anonymous referee for helpful comments that improved this work. We sincerely thank Tonghua Liu, Fan Yang and Bharat Ratra for their valuable suggestions and insightful discussions. %Kumar, D. thanks Bharat Ratra for useful discussions on the standardization of HII galaxies for use as distance indicators.  
% Zheng, J. was supported by National Natural Science Foundation of China under Grant No. 12403002; The Startup Research Fund of Henan Academy of Sciences (Project No. 241841221); The Scientific and Technological Research Project of Henan Academy of Science (Project No. 20252345001).
% Qiang, D.-C. was supported by the Startup Research Fund of Henan Academy of Sciences (Project No. 241841222).
% You, Z.-Q. was supported by the National Natural Science Foundation of China under Grant No. 12305059; The Startup Research Fund of Henan Academy of Sciences (Project No. 241841224); The Scientific and Technological Research Project of Henan Academy of Science (Project No. 20252345003); Joint Fund of Henan Province Science and Technology R\&D Program (Project No. 235200810111).  Kumar, D. is supported by the Startup Research Fund of the Henan Academy of Sciences under Grant number 241841219.        
   
% Bibliography    
   
%% [A] Recommended: using JHEP.bst file   
\bibliographystyle{JHEP} %cas-model2-names
\bibliography{main.bib}
   
%% or
%% [B] Manual formatting (see below)
%% (i) We suggest to always provide author, title and journal data or doi:
%% in short all the informations that clearly identify a document.
%% (ii) please avoid comments such as "For a review'', "For some examples",
%% "and references therein" or move them in the text. In general, please leave only references in the bibliography and move all
%% accessory text in footnotes.
%% (iii) Also, please have only one work for each \bibitem.

\end{document}